\documentclass[english,journal]{IEEEtran}
\usepackage[T1]{fontenc}
\usepackage[utf8]{inputenc}
\usepackage{color}
\usepackage{amsmath}
\usepackage{amssymb}
\usepackage{stmaryrd}
\usepackage{graphicx}
\usepackage{esint}
\usepackage{algorithm}
\usepackage{babel}

\begin{document}
\title{Gradient Descent Algorithm Inspired Adaptive Time Synchronization in Wireless Sensor Networks}

\author{Kasım~Sinan~Yıldırım,~\IEEEmembership{Member,~IEEE}
	
\thanks{Kasım~Sinan~Yıldırım is with Embedded Software Group, TU Delft, Delft, The Netherlands (k.s.yildirim@tudelft.nl).}
\thanks{}}


\maketitle
\begin{abstract}
Our motivation in this paper is to take another step forward from complex and heavyweight synchronization protocols to the easy-to-implement and lightweight synchronization protocols in WSNs. To this end, we present GraDeS, a novel multi-hop time synchronization protocol based upon gradient descent algorithm. We give details about our implementation of GraDeS and present its experimental evaluation in our testbed of MICAz sensor nodes. Our observations indicate that GraDeS is scalable, it has identical memory and processing overhead, better convergence time and comparable synchronization performance as compared to existing lightweight solutions.
\end{abstract}


\IEEEpeerreviewmaketitle

\section{Introduction}

\IEEEPARstart{R}{esource} constrained nodes in Wireless Sensor Networks (WSNs) are equipped with built-in hardware clocks that frequently drift apart due to their unstable low-cost crystal oscillators. Clock drift gives rise to loss of synchronization between sensor nodes which is problematic for collaborative and coordinated actions in WSNs, such as synchronous power on and shutdown of transceivers in order to reduce battery consumption. Therefore, each sensor node collects time information from its environment and constructs a software function, so-called logical clock, that represents network-wide global time. A logical clock is composed of an offset and a frequency that hold the value and speed difference between the corresponding hardware clock and the global time, respectively. The aim of time synchronization is to adjust the offsets and frequencies of the logical clocks so that estimation errors, i.e. clock skew, is minimized at any time instant.

There is an ample body of time synchronization protocols in the WSN literature, e.g. \cite{Maroti:2004:FTSP,Lenzen:2009:PulseSync,Yildirim:2014:Time-Synchronization-Based-on-Slow-Flooding-in-Wireless-Sensor-Networks}, that share the following steps to establish network-wide synchronization: (i) a reference node broadcasts its stable time information periodically; (ii) the receiver nodes adjust the offset and frequency of their logical clocks considering the received reference time; (iii) they broadcast the value of their logical clocks so that other nodes in the network synchronize by applying these steps. Least-squares regression is a common technique employed by these protocols to adjust the offsets and the frequencies of the logical clocks simultaneously \cite{Maroti:2004:FTSP,Lenzen:2009:PulseSync}. However, previous studies \cite{Yildirim:2014:Time-Synchronization-Based-on-Slow-Flooding-in-Wireless-Sensor-Networks,Yildirim:2014:Efficient-Time-Synchronization-in-a-Wireless-Sensor-Network-by-Adaptive-Value-Tracking} revealed that this technique is heavyweight in terms of processing and memory overhead, and simultaneous offset and frequency estimation gives rise to poor performance scalability. Other proposed techniques, such as maximum likelihood estimation \cite{Leng:11}, belief propagation \cite{Leng:2011} and convex closure \cite{Berthaud:00}, share the drawback of having heavy processing and memory overhead that allowed these studies to present only simulation results. Hence, the practicality of these alternative techniques is quite arguable. 

To overcome aforementioned problems in practice, two iterative methods are proposed \cite{Yildirim:2014:Efficient-Time-Synchronization-in-a-Wireless-Sensor-Network-by-Adaptive-Value-Tracking,PISync:2015}. AVTS protocol \cite{Yildirim:2014:Efficient-Time-Synchronization-in-a-Wireless-Sensor-Network-by-Adaptive-Value-Tracking} employs an efficient and adaptive search technique that adjusts the  relative frequency value of the clocks by observing the sign of the clock skew. The logical clock offset is adjusted independently from its frequency by adding the observed skew to the logical clock directly. On the other hand, PISync \cite{PISync:2015} is based upon a Proportional-Integral
(PI) controller and applies a proportional feedback (P) and an integral feedback (I) on the clock skew which allow to compensate offset and frequency differences, respectively. It has been reported that both iterative approaches achieve better and scalable synchronization performance with considerably less processing and memory overhead, and smaller code as compared to existing least-squares based protocols. As compared to PISync, AVTS requires knowledge about additional parameters in advance to perform its search effectively: the boundaries of the search space and the search precision affect its convergence speed and synchronization performance considerably. Hence, PISync appears to be a more promising solution in practice.

In this paper, our main contribution is to devise another iterative method to synchronize clocks in WSNs by introducing a novel time synchronization protocol, namely Gradient Descent Synchronization (GraDeS), which achieves scalable multi-hop time synchronization efficiently. In contrast to previous approaches, we formulate the frequency adjustment of the logical clocks as an optimization problem in which each sensor node is trying to find the frequency value of its logical clock that minimizes its synchronization error. Up to our knowledge of current literature, our study is the first to show that this optimization problem can be solved efficiently in practice by incorporating gradient descent algorithm \cite{Convex-Optimization-Boyd:2004}. We provide an extensive theoretical performance analysis of GraDeS as well as its practical implementation in TinyOS and evaluation in a testbed of 20 MICAz sensor nodes. Our theoretical and practical comparison with PISync revealed that both approaches exhibit nearly identical performances in terms of synchronization accuracy and resource overhead. As a brief conclusion, we believe that this study forms another step forward from complex and heavyweight to the easy-to-implement and lightweight time synchronization protocols in WSNs. 

The remainder of this paper is organized as follows: Next section presents the system model we will use in this paper. In section \ref{sec:A-New-Algorithm}, we present pairwise GraDeS protocol, its theoretical analysis and a comparison with PISync protocol. A multi-hop synchronization approach with GraDeS is presented in section \ref{sec:Mult-hop-GraDeS} and section \ref{sec:Experiments} gives details about implementation and evaluation in our testbed. Finally, section \ref{sec:Conclusion} is the conclusion.

\section{\label{sec:System-Model}System Model}

Our abstraction of a wireless sensor network is a \emph{connected}
graph $G=(V,E)$ whose vertex set $V=\left\{ 1,...,n\right\} $ represents
the identifiers of the sensor nodes and edge set $E\subseteq VxV$
represents the bidirectional communication links between these nodes.
Due to the \emph{broadcast} nature of wireless communication, once
a message is transmitted by any node $u\in V$, this message is received
by all nodes $v\in V$ such that $\left\{ u,v\right\} \in E$. We
refer these nodes, i.e. the nodes inside the communication range,
as the \emph{neighbors} of node $u$ and denote by $\mathcal{N}_{u}$. 

It is assumed sensor nodes are equipped with read-only hardware clocks
subject to clock drift. At any time $t>t_{0}$, we model the hardware
clock of any node $u$ as

\begin{eqnarray}
h_{u}(t) & = & h_{u}(t_{0})+\int_{t_{0}}^{t}f_{u}(\sigma)d\sigma
\end{eqnarray}
where $f_{u}(\sigma)\in[f_{0}-f_{max},f_{0}+f_{max}]$ denotes the
\emph{oscillator frequency} of the hardware clock, $f_{0}$ denotes
the \emph{nominal frequency} and $\pm f_{max}$ denotes the upper
and lower bounds of the\emph{ frequency deviation}. For the sake of
simplicity of the analytical steps in the following sections, we model the dynamic drift of the clocks as 

\begin{eqnarray}
f(t) & = & f_{0}+u(t)
\end{eqnarray}
such that $u(t)$ is a uniformly distributed random variable in the
interval $[-f_{max},f_{max}]$.

We assume that messages are never lost during communication. For any
message, the time that passes from the start of broadcast attempt
until the recipient node receives it is referred as \emph{transmission
delay}. Based on central limit theory and empirical observations \cite{Huang:2013:PSR:Practical-synchronous-rendezvous-in-low-duty-cycle-wireless-networks},
we model the transmission delays as a Gaussian distributed random
variable, denoted by $\mathcal{T}\sim\mathcal{N}(0,\sigma_{d}^{2})$.
The logical clock $l_{u}()$ of node $u$ can be modeled as

\begin{eqnarray}
l_{u}(t) & = & l_{u}(t_{up})+\hat{\Delta}_{u}(t_{up})(h_{u}(t)-h_{u}(t_{up}))
\end{eqnarray}
where $t_{up}$ is the latest time at which the logical clock is updated.
In this model, \emph{rate multiplier} $\hat{\Delta}_{u}(t_{up})$
is the estimate of the relative frequency $f_{0}/f_{u}(t)$ in the
interval $[t_{0},t]$ and it is modified to speed-up or slow-down
the logical clock. The \emph{offset} $l_{u}(t_{up})$ is used to correct
the value of the logical clock with fresh time information. According
this model, time synchronization can simply be considered as a distributed
algorithm which updates the logical clock parameters $l_{u}(t_{up})$
and $\hat{\Delta}_{u}(t_{up})$ of each node $u$ at each update time
$t_{up}$.

\section{\label{sec:A-New-Algorithm}A New Time Synchronization Algorithm
for Wireless Sensor Networks: GraDeS}

In unconstrained optimization problems, the objective is to minimize
a function $f(x)$ where $f:\mathbb{R}^{n}\rightarrow\mathbb{R}$
is convex and differentiable. It is assumed that the problem is solvable
and hence there is an optimal point $x^{*}$. Since $f$ is convex
and differentiable, $\nabla f(x^{*})=0$. The \emph{descent methods
}produce a sequence $x_{k+1}=x_{k}+\alpha_{k}\Delta x_{k}$ such that
$f(x_{k+1})<f(x_{k})$ where $k=0,1,...$ denotes the\emph{ iteration
number}, $\alpha_{k}>0$ is the \emph{step size} and $\Delta x_{k}$
is the descent direction at iteration $k$. Starting from an initial
point $x_{0}$, a descent direction is determined and a step size
is chosen at the beginning of each iteration to obtain the new sequence
value. This iteration is continued until convergence. When the search
direction is determined as the negative gradient $\Delta x=-\nabla f(x)$,
the resulting algorithm is called the \emph{gradient descent algorithm}
\cite{Convex-Optimization-Boyd:2004}. In the following subsections,
we introduce a new time synchronization protocol, namely Gradient
Descent Synchronization (GraDeS), which is inspired from this algorithm.

\subsection{Pairwise Synchronization with GraDeS Approach}

Consider pairwise time synchronization of two sensor nodes $u$ and
$r$, where $r$ is the reference node which has access to the real-time
$t$. Assume that node $r$ transmits messages with a period of $B$
seconds in order to inform node $u$ about $t$. Let $t_{h}=Bh$ for
$h=0,1,...$, be the packet reception times of node $u$ from node
$r$ and let $l_{r}(t_{h})=B.h+\mathcal{T}_{h}$ be the received clock
where $\mathcal{T}_{h}$ denotes the transmission delay at the step
\emph{h}. The synchronization error of node \emph{u} with respect
to the reference node\textcolor{blue}{{} }$r$ at any packet reception
time $t_{h}=Bh$ can be calculated as

\begin{eqnarray}
e_{u}(t_{h}) & \!\!=\!\! & l_{u}(t_{h})-l_{r}(t_{h})-\mathcal{T}_{h}=l_{u}(t_{h})-Bh-\mathcal{T}_{h}.
\end{eqnarray}
Let node \emph{u} simply sets its logical clock to the received clock
value to compensate the offset difference between the clocks. Formally,
assume that node \emph{u} applies the following correction to its
logical clock 

\begin{eqnarray}
l_{u}(t_{h}^{+}) & = & l_{u}(t_{h})-e_{u}(t_{h})
\end{eqnarray}
where $t_{h}^{+}$ denotes the time instant just after $t_{h}$. After
this compensation, the \emph{synchronization error} $e_{u}(t_{h+1})$
at the subsequent packet reception time $t_{h+1}$ will be mainly
due to the different hardware clock frequency of the $u$. Applying
straightforward steps, the function $e_{u}()$ can be generalized
as:

\begin{eqnarray}
e_{u}(t_{h+1}) & \!\!\!\!\!= & \!\!\!\!\!\hat{\Delta}_{u}(t_{h}^{+})\int_{t_{h}^{+}}^{t_{h+1}}f_{u}(t)dt-(B+\mathcal{T}_{h+1}-\mathcal{T}_{h})\label{eq:error_evolution}
\end{eqnarray}
where $e_{u}(t_{0})=0$ and $\hat{\Delta}_{u}(t_{0})=1$.

The objective of time synchronization is to minimize the synchronization
error which, in our case, is the\textcolor{blue}{{} }function $e_{u}()$
of the dynamic parameter $\hat{\Delta}_{u}$. Getting inspired from
the \emph{gradient descent algorithm}, our objective reduces into
finding the optimal value of $\hat{\Delta}_{u}^{*}$ in order to minimize
the squared error function $(e_{u}())^{2}$. Formally, we define the
steps of the GraDeS algorithm that will be employed at at each updating
time instant $t_{h}$ as follows: 

\begin{eqnarray}
l_{u}(t_{h}^{+}) & = & t_{h}+\mathcal{T}_{h},\\
\hat{\Delta}_{u}(t_{h}^{+}) & = & \hat{\Delta}_{u}(t_{h})-\alpha\nabla e_{u}(t_{h})\label{eq:delta}
\end{eqnarray}
It should be noted that for all $t\in[t_{h}^{+},t_{h+1}^{+})$ and
$\hat{\Delta}_{u}(t_{h})=\hat{\Delta}_{u}(t_{h}^{+})$. In the update
equation above, $\nabla e_{u}(t_{h})$ denotes the derivative of $(e_{u}(t_{h}))^{2}$
with respect to $\hat{\Delta}_{u}(t_{h})$ at time $t_{h}$. From
the equation \eqref{eq:error_evolution}, it can be observed that
the function $(e_{u}(t_{h}))^{2}$ is continuous and differentiable
in the interval $[t_{h-1},t_{h}]$. Therefore,

\begin{eqnarray}
\nabla e_{u}(t_{h}) & = & 2e_{u}(t_{h})\frac{de_{u}(\hat{\Delta}_{u})}{d\hat{\Delta}_{u}}.
\end{eqnarray}

\subsection{Approximation of the Error Derivative}

The derivative of $\frac{de_{u}(\hat{\Delta}_{u})}{d\hat{\Delta}_{u}}$
in the interval $\left[t_{h},t_{h+1}\right]$ can be approximated
with the following equation:

\begin{eqnarray}
\frac{de_{u}(\hat{\Delta}_{u})}{d\hat{\Delta}_{u}} & = & \frac{d\left(\hat{\Delta}_{u}\int_{t_{h}^{+}}^{t_{h+1}}f_{u}(t)dt-(B+\mathcal{T}_{h+1}-\mathcal{T}_{h})\right)}{d\hat{\Delta}_{u}}\nonumber \\
 & = & \int_{t_{h}^{+}}^{t_{h+1}}f_{u}(t)dt.
\end{eqnarray}
We have $E\left\llbracket \int_{t_{h}}^{t_{h+1}}f_{u}(t)dt\right\rrbracket =Bf_{0}$
due to the fact that the frequency drift of the clocks are modeled
as uniform random variables in Section \eqref{sec:System-Model}.
Therefore, the derivative becomes $\frac{de_{u}(t_{h})}{d\hat{\Delta}_{u}}=Bf_{0}$
in expectation. Finally,\textcolor{blue}{{} }we get

\begin{eqnarray}
\nabla e_{u}(t_{h}) & = & 2Bf_{0}e_{u}(t_{h}).
\end{eqnarray}

\subsection{Proof of Convergence and Steady-State Error}

With an abuse of notation, let us denote $e(t_{h+1})$ by $e(h+1)$
and $\hat{\Delta}_{u}(t_{h+1})$ by $\hat{\Delta}_{u}(h+1)$. Based
upon the update equations of GraDeS algorithm, the system evolution
can be described with the following matrix equation:

\begin{eqnarray}
\underset{X(h+1)}{\underbrace{\left[\!\!\!\begin{array}{c}
e(h+1)\\
\hat{\Delta}_{u}(h+1)
\end{array}\!\!\!\right]}} & \!\!\!\!\!\!\!\!=\!\!\!\!\!\!\!\! & \left[\!\!\!\begin{array}{cc}
0 & \int_{t_{h}}^{t_{h+1}}f_{u}(t)dt\\
0 & 1-2\alpha Bf_{0}\int_{t_{h}}^{t_{h+1}}f_{u}(t)dt
\end{array}\!\!\!\right]\underset{X(h)}{\underbrace{\left[\begin{array}{c}
e(h)\\
\hat{\Delta}_{u}(h)
\end{array}\right]}}\nonumber \\
 &  & +(B\!+\!\mathcal{T}_{h+1}\!-\!\mathcal{T}_{h}\!)\!\!\left[\!\!\!\!\begin{array}{c}
-1\\
2\alpha Bf_{0}
\end{array}\!\!\!\!\right].\label{eq:state_space}
\end{eqnarray}
Taking the expectation of both sides yields:

\begin{eqnarray}
E\!\left\llbracket X(h+1)\right\rrbracket \!\!\!\!\!\! & = & \!\!\!\!\!\!\!\underset{A}{\underbrace{\left[\!\!\!\begin{array}{cc}
0 & Bf_{0}\\
0 & \!\!\!\!\!1-2\alpha B^{2}f_{0}^{2}
\end{array}\!\!\!\!\right]}\!}E\!\left\llbracket X(h)\right\rrbracket +\!\!\left[\!\!\!\!\begin{array}{c}
-B\\
2\alpha B^{2}f_{0}
\end{array}\!\!\!\!\right].
\end{eqnarray}
The eigenvalues of the matrix $A$ can be calculated by solving the
determinant equation $|A-\lambda I|=0$. The solution of this equation
can be obtained by solving the following quadratic equation

\begin{eqnarray}
\lambda\left(1-2\alpha B^{2}f_{0}^{2}-\lambda\right) & = & 0
\end{eqnarray}
whose roots are:

\begin{align}
\lambda_{1}=0, & \lambda_{2}=1-2\alpha B^{2}f_{0}^{2}.\label{eq:roots}
\end{align}
Matrix $A$ is asymptotically stable, i.e. asymptotic convergence
is established, if and only if $|\lambda_{1,2}|<1$. Therefore, choosing
the step size by considering the inequality below

\begin{align}
0 & <\alpha<\frac{1}{B^{2}f_{0}^{2}}\label{eq:step_size_bounds}
\end{align}
will lead the system to converge to the asymptotically stable equilibrium
point 
\begin{eqnarray}
\left[\underset{h\rightarrow\infty}{lim}E\left\llbracket e(h)\right\rrbracket ,\underset{h\rightarrow\infty}{lim}E\left\llbracket \hat{\Delta}_{u}(h)\right\rrbracket \right]^{T} & \!\!\!\!\!\!\!\!\!\!\!=\!\!\!\!\!\!\! & \left[e(\infty),\hat{\Delta}_{u}(\infty)\right]^{T}.
\end{eqnarray}
Therefore

\begin{eqnarray*}
\hat{\Delta}_{u}(\infty) & \!\!\!\!\!\!=\!\!\!\!\!\! & \left(1-2\alpha B^{2}f_{0}^{2}\right)\hat{\Delta}_{u}(\infty)+2\alpha B^{2}f_{0}.
\end{eqnarray*}
which indicates that 
\begin{eqnarray}
\hat{\Delta}_{u}(\infty) & = & \frac{1}{f_{0}}.\label{eq:mean_drift_error}
\end{eqnarray}
Similarly,

\begin{eqnarray}
e_{u}(\infty) & = & B(\hat{\Delta}_{u}(\infty)f_{0}-1)=0.\label{eq:mean_offset_error}
\end{eqnarray}
The expressions above show that time synchronization will eventually
be achieved with an expected steady-state error of $e_{u}(\infty)=0$. 

In Appendix-\eqref{appendix:Asymptotic-Variance}, we show that the
variance of this approach can be calculated as 

\begin{eqnarray}
Var(e(\infty)) & \!\!\!\!\!=\!\!\!\!\! & \frac{\alpha\!\left(\!\!\frac{Bf_{max}^{2}}{3}\!+\!f_{0}^{2}\sigma_{d}^{2}\!\right)}{1\!\!-\!\alpha\!\left(\!\!B^{2}f_{0}^{2}\!+\!\frac{Bf_{max}^{2}}{3}\!\!\right)}\!\!\left(\!\!B^{2}\!+\!\frac{Bf_{max}^{2}}{3f_{0}^{2}}\!\!\right)\nonumber \\
 &  & +\frac{Bf_{max}^{2}}{3f_{0}^{2}}+\!\sigma_{d}^{2}.\label{eq:GraDeS_variance}
\end{eqnarray}
It is apparent that as long as the inequality \eqref{eq:step_size_bounds}
is satisfied, the convergence will be established. However, the smaller
the value of $\alpha$, the smaller the asymptotic variance. On the
other hand, the convergence time is inversely related by the magnitude
of the eigenvalue $\lambda_{2}$ in equality \eqref{eq:roots}. Hence,
the smaller the $\alpha$, the bigger $\lambda_{2}$ is and thus the
longer the convergence time. 

\begin{figure}
\center

\includegraphics[scale=0.45]{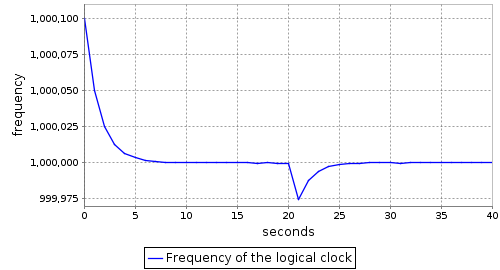}

\caption{\label{fig:Gradient-descent-simulations}Simulation results related
to $f_{u}\hat{\Delta}_{u}$ during the synchronization between the
reference node $r$ and the node u with the following system parameters:
$B=30$ seconds, $f_{0}=f_{r}=1MHz$, $f_{max}=100\,ppm$ and $\mathcal{T}\sim\mathcal{N}(0,\sigma_{d}^{2}=100\,microseconds)$.
We set $f_{u}-f_{r}=100$ ppm and $\alpha_{u}=0.5$. After a finite
number of iterations, the logical clock frequency of node $u$ converges
to $f_{r}$. At iteration 20, the hardware clock frequency of node
$u$ is modified by setting $f_{u}-f_{r}=50$ ppm. In this case, the
logical clock frequency of node $u$ gets adapted and converges to
$f_{r}$ again. }
\end{figure}

Figure \eqref{fig:Gradient-descent-simulations} presents the evolution
of the frequency of the logical clock of node $u$ during a numerical
simulation, from which we observe that the proposed synchronization
approach establishes synchronization between two nodes and it is adaptive
in terms of environmental dynamics, that fits perfectly for the wireless
sensor networks.

\subsection{Theoretical Comparison to PISync Algorithm}

In Appendix-\eqref{appendix:Comparison}, we anaylzed
PI-Controller based PISync algorithm \cite{PISync:2015} under the
same system model. Based on these results, we summarize the differences and similarities 
between GraDeS and PISync as follows:
\begin{itemize}
\item Since the largest eigenvector of the system matrix
A of GraDeS is smaller than that of PISync, i.e. $\lambda_{2}^{GraDeS}=1-2\alpha B^{2}f_{0}^{2}<\lambda_{2}^{PISync}=1-2\alpha Bf_{0}$
when $\alpha$ parameters are identical, time-to-convergence of GraDeS
is superior than that of PISync.
\item  Considering asymptotic variances of GraDeS in equality
\eqref{eq:GraDeS_variance} and that of PISync in equality \eqref{eq:pisync_variance},
as long as the re-synchronization period satisfies $B<\frac{1}{2f_{0}}$
and $\alpha$ parameters are identical, the synchronization performance
of GraDeS is superior than that of PISync. However, when practical
parameters in Figure \eqref{fig:Gradient-descent-simulations} are
taken into account, such a synchronization period is not applicable
which makes PISync a better choice, in theory.

\item Since the update equations of GraDeS and PISync are quite similar, errors introduced by communication delays, quantization, etc., enter linearly to the system equations.\footnote{Details are already given in \cite{PISync:2015}}. Therefore, the synchronization errors of both approaches grow with the \textit{square-root} of the network diameter, that gives rise to scalable synchronization performance degradation. 
\end{itemize}

\section{\label{sec:Mult-hop-GraDeS} Multi-hop Synchronization of WSNs with GraDeS}

\begin{algorithm}
	\caption{\label{alg:Multi-Hop}GraDeS pseudo-code for node $u$.}
	
	1: \textbf{Upon receiving} $\left\langle {l_{v}},seq_{v}\right\rangle $ \textbf{such that} $seq_{v}>seq_{u}$  
	
	2: \: $seq{}_{u}\leftarrow seq_{v}$
	
	3: \: $e{}_{u}\leftarrow{l_{u}-l_{v}}$
	
	4: \: $l_{u}\leftarrow l_{v}$
	
	5: \: \textbf{update}
			$\alpha_{u}$
	
	6: \: $\hat{\Delta}_{u}\leftarrow\hat{\Delta}_{u}-\alpha_u\nabla e_{u}$
	
	7:
	
	8: \textbf{Upon} $h_u=kB$ \textbf{where} $k \in \mathbb{N}$  
	
	9: \: \textbf{if} $u=r$ \textbf{then} $seq_{u}\leftarrow seq_{u}+1$
	
	10: \textbf{broadcast} $\left\langle {l_{u}},seq_{u}\right\rangle $
\end{algorithm}

Algorithm \eqref{alg:Multi-Hop} presents the pseudo-code for the
Gradient Descent Synchronization (GraDeS) protocol that extends our
pairwise synchronization scheme to multi-hop. For simplicity, it is
assumed that the reference node $r$ is predefined before the deployment
of the sensor network. However, simple root election mechanisms (e.g.
in \cite{Maroti:2004:FTSP}) can easily be integrated to the protocol.
Whenever the hardware clock of any node reaches a multiple of $B$
(Line 8), only the reference node increments the sequence number (Line
9). Then, each node broadcasts a synchronization message that carries
the value of its logical clock and its sequence number for their neighboring
nodes in order to establish network-wide time synchronization (Line
10). It should be noted that $l_{r}=h_{r}$ for
the reference node and it does not participate in the synchronization
process.

Sensor nodes other than the reference collect the synchronization
messages that belong to the new synchronization round, i.e. with higher
sequence numbers (Line 1). At the first step, they update their sequence
number (Line 2) and calculate the synchronization error (Line 3).
Then, they set their logical clock to the received time information
for offset compensation (Line 4). Following this step, they update their step sizes (Line 5). We will explain the details of this step in the following paragraphs. Finally, they update their
rate multipliers according to the gradient descent algorithm (Line
6). Observe that, in contrast to the regression table in least-squares,
nodes executing GraDeS protocol do not require any memory to collect
time information of the reference node. The operations during logical
clock update (Lines 3-6) are quite simple and easy to implement as
compared to the calculation of the least-squares line. 
\subsection{Adaptation of the Step Size}
The step size $\alpha$ has an important effect on both synchronization error performance and convergence time of the GraDeS
algorithm. Choosing a constant and big step size would lead to a faster
convergence but also a big steady state synchronization error. On
the other hand, choosing a constant and small step size would lead
to a slow convergence but smaller steady state synchronization error.
When environmental dynamics in WSNs are considered, individual sensors
should react to these changes fast and slow convergence would lead
to a big problem. For this purpose, we 
modified the adaptation algorithm in \cite{PISync:2015}, that adjusts step size adaptively in order to achieve fast convergence and small steady-state error, as shown below:

\begin{eqnarray}
\alpha(t_{h}) & = & \begin{cases}
{2\alpha(t_{h-1})} & \mbox{{\bf if} }\nabla e(t_{h}).\nabla e(t_{h-1})\!>\!0\\
{1/3\alpha(t_{h-1})} & \mbox{{\bf otherwise} }
\end{cases}.\label{eq:adaptation}
\end{eqnarray}

\begin{figure}
	\centering
	\includegraphics[scale=0.45]{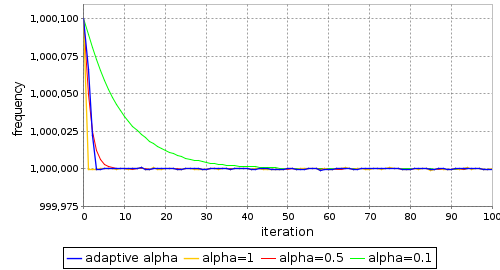}
	
	\caption{\label{fig:Gradient-descent-adaptive_simulations}Simulation results related to the evolution of the logical clock frequency with different step sizes during the synchronization and between the reference node $r$ and the node u. The parameter setting in Figure  \eqref{fig:Gradient-descent-simulations} were also used for these simulations. With constant step size, the convergence time increases but the steady state error decreases as the step size gets smaller. With adaptive algorithm, the convergence time is faster and the steady state error is quite comparable to those with constant step sizes.}
\end{figure}

The intuition behind this approach is similar to that presented in \cite{PISync:2015}: Let $t_{h}$ be the receipt time of a new synchronization message and let $\nabla e(t_{h})$ be the derivative of the error observed 	at that time. If the derivative $\nabla e(t_{h})$ and the derivative
of the previous round $\nabla e(t_{h-1})$ have the same sign, i.e.
their directions are the same, then the more $\hat{\Delta}$ is supposed
to be far away from its optimal value $\hat{\Delta}^{*}$. Hence,
it necessary to \textit{accelerate} the adjustment of $\hat{\Delta}$
in order to reach $\hat{\Delta}^{*}$ more quickly. On the contrary,
if the signs of $\nabla e(t_{h})$ and $\nabla e(t_{h-1})$ are opposite,
then $\hat{\Delta}$ is oscillating around $\hat{\Delta}^{*}$. In
order to get closer to the optimal value, it is necessary to \textit{decelerate}
the adjustment. It is worth to mention that this algorithm is inspired
from \cite{Yildirim:2014:Efficient-Time-Synchronization-in-a-Wireless-Sensor-Network-by-Adaptive-Value-Tracking}
in which the multipliers $2$ and $1/3$ are shown
to be good values in terms of convergence performance. Figure \eqref{fig:Gradient-descent-adaptive_simulations}
presents the evolution of the logical clock frequency of node $u$
with constant and adaptive step sizes during a numerical simulation.\footnote{It should be noted that if $\alpha(t_{h})>\frac{1}{B^2f_0^2}$ then $\alpha(t_{h})=\frac{1}{B^2f_0^2}$
	and if $\alpha(t_{h})=0$ then $\alpha(t_{h})=\alpha(t_{h-1})$.}

\begin{figure*}
	\center
	\includegraphics[scale=0.28]{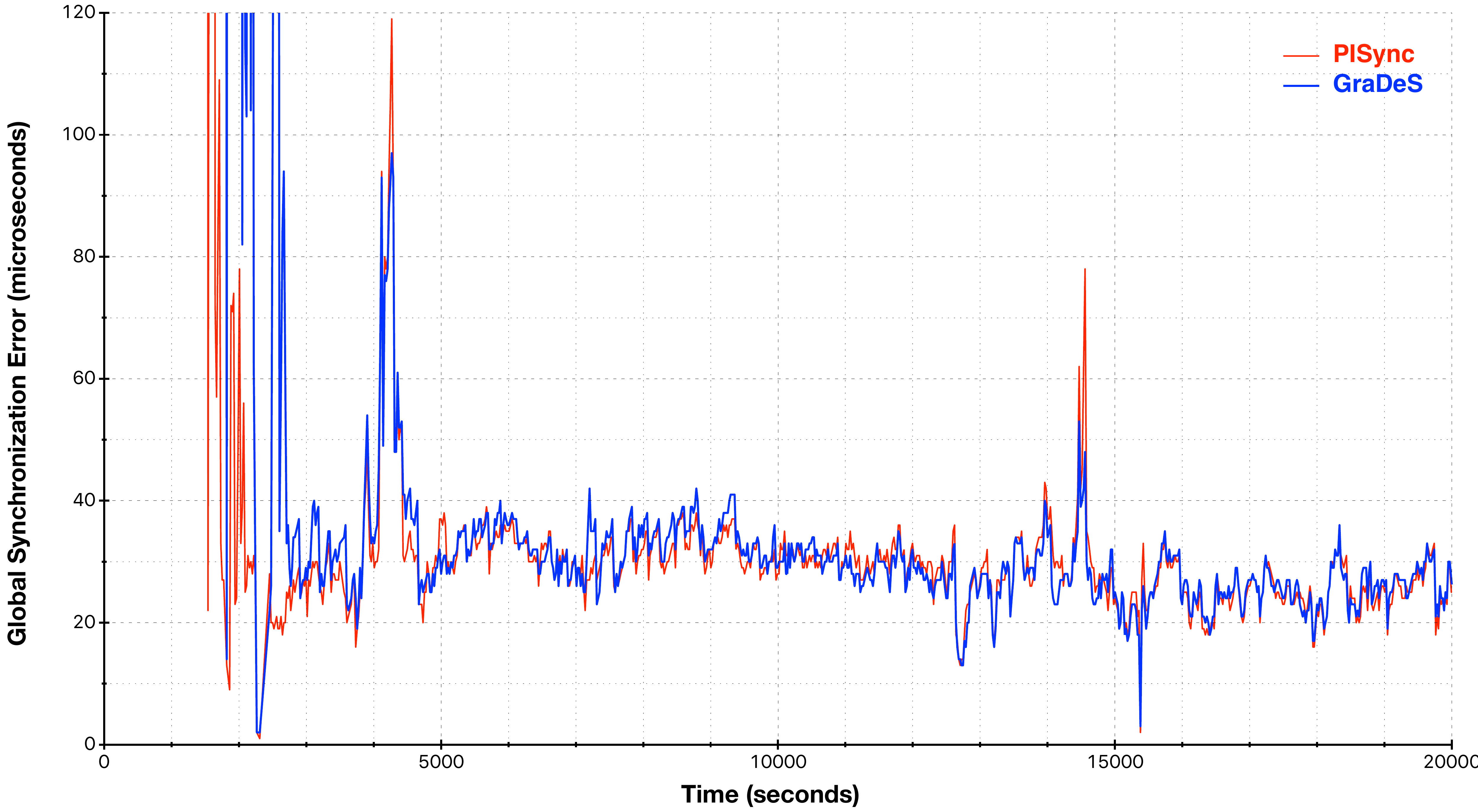}
	
	\caption{\label{fig:Global-Skew}Instantaneous and average global synchronization
		errors for GraDeS and PISync.}
\end{figure*}

\section{\label{sec:Experiments} Experimental Evaluation} 

We implemented GraDeS and PISync protocols for MICAz sensor platform
using TinyOS 2.1.2 operating system. In order to 
evaluate their multi-hop synchronization performances, we prepared identical
testbed setup as presented in \cite{Lenzen:2009:PulseSync,Yildirim:2014:Time-Synchronization-Based-on-Slow-Flooding-in-Wireless-Sensor-Networks,Yildirim:2014:Efficient-Time-Synchronization-in-a-Wireless-Sensor-Network-by-Adaptive-Value-Tracking} and preferred a line topology of 20 sensor nodes to evaluate scalability and adaptivity. We used 7.37 MHz quartz oscillator on the MICAz board as the clock source for the timer used for timing measurements. The timer operates at 1/8 of that frequency and thus each timer
tick occurs at approximately every 921 kHz, i.e., approximately 1 microseconds, therefore, $f_0=1$ MHz.
During the experiments, each lasted approximately 20000 seconds, we fixed
beacon period $B=30$ seconds and we collected instantaneous
logical clock values from the nodes. For performance
comparison, we considered \textit{global skew} which is defined as the largest instantaneous clock difference between arbitrary nodes.

Performance evaluation of synchronization protocols in a fair manner is a challanging task since it is almost impossible to create identical message delays, packet loss rates and environmental conditions during tests that effect the frequency of the crystal oscillators of the sensor nodes. To this end, we followed the same approach in \cite{Yildirim:2012:Drift-Estimation-Using-Pairwise-Slope-with-Minimum-Variance-in-Wireless-Sensor-Networks} and we integrated both protocols to each sensor node since both GraDeS and PISync have identical message patterns. First, we enlarged the synchronization messages so that they carry the logical clock values calculated both using GraDeS and PISync.  In this manner, when a synchronization message is received by a sensor
node, it extracts the logical clock value for GraDeS and that for PISync to update its corresponding logical clocks. As a final modification,
we added an interface to
query the logical clock values calculated with GraDeS and PISync. With such modifications, we could evaluate both strategies under identical executions.

Figure \eqref{fig:Global-Skew} presents the synchronization
performances of PISync and GraDeS. First, we observed that the convergence
times of GraDeS and PISync were almost identical. Even though our theoretical comparison provided a slightly superior convergence time for GraDeS, this superiority was not observable in practice due to the practical values of the system parameters. Our second observation
is the superiority of the synchronization performance of GraDeS over
PISync at the time instants where there were error peaks after convergence is established, i.e. the peak around second 4300 and that around 14600. Apart from these points, the performances of both approaches were quite comparable. The reason for this phenomenon is related to the step size $\alpha$ values. It is apparent that the step size boundaries of GraDeS, i.e. $(0,\frac{1}{B^2f_0^2}]$ is quite narrow than that of PISync, i.e. $(0,\frac{2}{Bf_0}]$. This led to smaller step size values for GraDeS that allowed to be more robust against erroneous nodes. Moreover, step size adaptation of GraDeS is different than that of PISync, since in the former considers the sign of the derivative of the error whereas the latter considers the sign of the error. These differences led to a different reactive behavior. During the experiments, maximum global skew values after the convergence
were 96 and 119 microseconds for GraDeS and PISync, respectively. However, neglecting peak points, their performances were quite identical. 

For efficient duty-cycling of the radios, nodes should
	estimate when data is coming to switch on their radios for receiving
	the data. In particular, a \emph{guard time}
	is necessary to compensate the synchronization errors. As indicated
	in \cite{OPTIMAL_SLEEP:2009}, existing sleep/wake scheduling schemes
	assume that the underlying synchronization protocol can provide microsecond-level
	synchronization so that their guard times are small and nodes keep
	their radios on for a less amount of time, leading to less energy
	consumption. Therefore, the microsecond synchronization performance
	of GraDeS meets the typical requirements of the existing duty-cycling
	schemes and it can effectively be used by them. In our implementations, PISync and GraDeS had identical
	main memory overhead since they maintain three 32 bit variables $\hat{t}_{u}(t_{up})$,
	$\hat{\Delta}_{u}(t_{up})$ and $h_{u}(t_{up})$ for logical clock,
	and additional 32 bits are required for the step size adaptation. Since these protocols
	have identical communication frequencies, their energy consumption
	during a packet processing and updating the logical clock are quite comparable. We refer the reader \cite{PISync:Arxiv} for a detailed discussion of these issues.

\section{\label{sec:Conclusion}Conclusion and Future Work}

In this article, we formulated pairwise synchronization process as an optimization
problem and showed that it can efficiently be solved by employing
gradient descent algorithm. We introduced a new time synchronization
protocol, namely \emph{Gradient Descent Synchronization (GraDeS)},
that establishes multi-hop synchronization based upon this algorithm.
We gave details about our implementation and presented the experimental
evaluation on our testbed of MICAz sensor nodes. A future research direction can be the integration of the proposed approach to duty-cycling MAC protocols in WSNs in order to observe its impact on the conservation of the energy. Another point worth to explore is to integrate GraDeS to real-world applications to evaluate its actual performance. 

\bibliographystyle{IEEEtran}
\bibliography{references}

\begin{IEEEbiography}[{\includegraphics[width=1in,height=1.25in,clip,keepaspectratio]{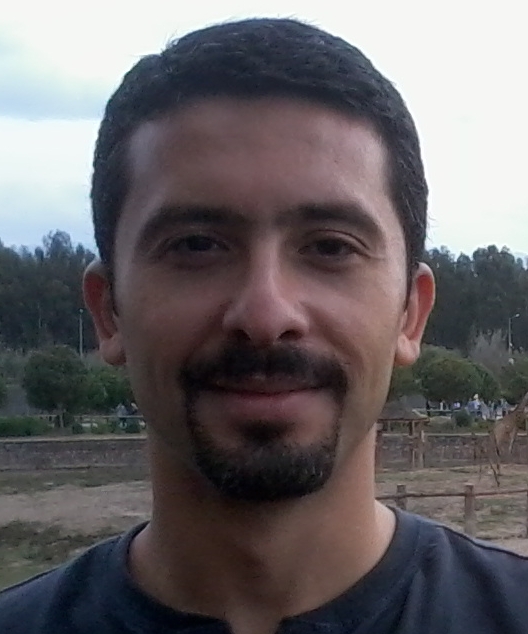}}]
	{Kasım Sinan Yıldırım} 	received the BSc(Eng), M.Sc. and Ph.D. degrees from Department of Computer Engineering, Ege University, İzmir, Turkey in 2003, 2006 and 2012 respectively. He worked as a research assistant between 2007-2013 and as an assistant professor between 2013-2015 at the same department. Currently, he is a postdoctoral researcher at Embedded Software Group, TU Delft, The Netherlands. His research interests are embedded systems, distributed systems, distributed algorithms,  wireless sensor networks, wireless power transfer networks and computational RFIDs. He is a member of IEEE. 
\end{IEEEbiography}

\appendices

\section{Asymptotic Variance of GraDeS\label{appendix:Asymptotic-Variance}}

Define $w_{h+1}=\int_{_{t_{h}}}^{t_{h+1}}u(t)dt$. Applying straightforward
steps, $\int_{_{h^{+}}}^{t_{h+1}}f(t)dt$ becomes:

\begin{eqnarray}
\int_{_{h^{+}}}^{t_{h+1}}f(t)dt & = & \int_{_{h^{+}}}^{t_{h+1}}\left(f_{0}+u(t)\right)dt\nonumber \\
 & = & Bf_{0}+\int_{_{t_{h}}}^{t_{h+1}}u(t)dt\nonumber \\
 & = & Bf_{0}+w_{h+1}
\end{eqnarray}
From the definition of $w_{h+1}$ and the properties of the uniform
random variables, it holds that $E\left\llbracket w_{h+1}\right\rrbracket =0$
and $E\left\llbracket w_{h+1}^{2}\right\rrbracket =\frac{Bf_{max}^{2}}{3}$.
Now, define $z_{h}=\hat{\Delta}(h)f_{0}-1$ and $d_{h+1}=\mathcal{T}_{h+1}-\mathcal{T}_{h}$.
With these definitions, the error function can be written as:

\begin{eqnarray}
e(h+1) & = & \hat{\Delta}(h)\int_{h^{+}}^{h+1}f(t)dt-(B+\mathcal{T}_{h+1}-\mathcal{T}_{h})\nonumber \\
 & = & Bz_{h}+\hat{\Delta}(h)w_{h+1}-d_{h+1}\nonumber \\
 & = & Bz_{h}+\left(\frac{z_{h}+1}{f_{0}}\right)w_{h+1}-d_{h+1}\nonumber \\
 & = & z_{h}\left(B+\frac{w_{h+1}}{f_{0}}\right)+\frac{w_{h+1}}{f_{0}}-d_{h+1}.\label{eq:error_zh}
\end{eqnarray}
Similarly, for $\hat{\Delta}(h+1)$ we can apply the following steps:

\begin{eqnarray}
\hat{\Delta}(h+1) & = & \hat{\Delta}(h)-2\alpha Bf_{0}e(h+1)\nonumber \\
 & = & \frac{z_{h}+1}{f_{0}}\nonumber \\
 &  & \!\!\!\!\!\!-2\alpha Bf_{0}\!\left(\!\!z_{h}\!\!\left(\!\!B\!\!+\!\frac{w_{h+1}}{f_{0}}\!\!\right)\!\!+\!\frac{w_{h+1}}{f_{0}}\!\!-\!d_{h+1}\!\!\right)\nonumber \\
 & = & z_{h}\left(\frac{1}{f_{0}}-2\alpha Bf_{0}\left(B+\frac{w_{h+1}}{f_{0}}\right)\right)\nonumber \\
 &  & +\frac{1}{f_{0}}-2\alpha Bf_{0}\left(\!\!\frac{w_{h+1}}{f_{0}}\!\!-\!d_{h+1}\!\!\right).
\end{eqnarray}
Let $g_{h+1}=Bf_{0}+w_{h+1}$. Then, $z_{h+1}=\hat{\Delta}(h+1)f_{0}-1$
becomes:

\begin{eqnarray}
z_{h+1} & = & z_{h}\left(1-2\alpha Bf_{0}g_{h+1}\right)-2\alpha Bf_{0}g_{h+1}\nonumber \\
 &  & +2\alpha Bf_{0}^{2}(B+d_{h+1}).
\end{eqnarray}
It easy to see that $z_{h}$, $g_{h+1}$ and $d_{h+1}$ are independent
random variables. Similarly, we have $E\left\llbracket g_{h+1}\right\rrbracket =Bf_{0}$
and $E\left\llbracket g_{h+1}^{2}\right\rrbracket =B^{2}f_{0}^{2}+\frac{Bf_{max}^{2}}{3}$.
Therefore:

\begin{eqnarray}
E\left\llbracket z_{h+1}\right\rrbracket  & = & E\left\llbracket z_{h}\right\rrbracket (1-2\alpha B^{2}f_{0}^{2}).
\end{eqnarray}
that clearly shows that 
\begin{eqnarray}
\underset{h\rightarrow\infty}{lim}E\left\llbracket z_{h}\right\rrbracket  & = & 0.
\end{eqnarray}
Similarly, after some calculations, we can obtain:

\begin{eqnarray*}
\underset{h\rightarrow\infty}{lim}E\left\llbracket z_{h}^{2}\right\rrbracket  & = & \frac{\alpha\left(\frac{Bf_{max}^{2}}{3}+f_{0}^{2}\sigma_{d}^{2}\right)}{1-\alpha\left(B^{2}f_{0}^{2}+\frac{Bf_{max}^{2}}{3}\right)}.
\end{eqnarray*}

Now, let us bring our attention to the asymptotic variance of the
error function. Considering equation \eqref{eq:error_zh}, it holds
that 

\begin{eqnarray}
\underset{h\rightarrow\infty}{lim}E\left\llbracket e(h+1)\right\rrbracket  & = & BE\left\llbracket z_{h}\right\rrbracket =0.
\end{eqnarray}
and

\begin{eqnarray}
\underset{h\rightarrow\infty}{lim}E\left\llbracket \left(e(h+1)\right)^{2}\right\rrbracket  & = & E\left\llbracket z_{h}^{2}\right\rrbracket \left(B^{2}+\frac{E\left\llbracket w_{h+1}^{2}\right\rrbracket }{f_{0}^{2}}\right)\nonumber \\
 &  & +\frac{E\left\llbracket w_{h+1}^{2}\right\rrbracket }{f_{0}^{2}}+E\left\llbracket d_{h+1}^{2}\right\rrbracket \nonumber \\
 & =\!\!\!\!\!\!\!\! & \frac{\alpha\!\left(\!\!\frac{Bf_{max}^{2}}{3}\!+\!f_{0}^{2}\sigma_{d}^{2}\!\right)}{1\!\!-\!\alpha\!\left(\!\!B^{2}f_{0}^{2}\!+\!\frac{Bf_{max}^{2}}{3}\!\!\right)}\!\!\left(\!\!B^{2}\!+\!\frac{Bf_{max}^{2}}{3f_{0}^{2}}\!\!\right)\!\nonumber \\
 &  & +\frac{Bf_{max}^{2}}{3f_{0}^{2}}+\!\sigma_{d}^{2}.\label{eq:grades_variance}
\end{eqnarray}
Finally, the asymptotic error variance can be obtained as follows:

\begin{eqnarray}
\underset{h\rightarrow\infty}{lim}Var(e(h+1)) & = & E\left\llbracket \left(e(h+1)\right)^{2}\right\rrbracket -\left(E\left\llbracket e(h+1)\right\rrbracket \right)^{2}\nonumber \\
 & = & E\left\llbracket \left(e(h+1)\right)^{2}\right\rrbracket .
\end{eqnarray}

\section{Theoretical Comparison to PISync\label{appendix:Comparison}}

In \cite{PISync:2015}, the update equations of the PI controller
based time synchronization protocol PISync is given as:

\begin{eqnarray}
l_{u}(t_{h}^{+}) & = & t_{h}+\mathcal{T}_{h},\\
\hat{\Delta}_{u}(t_{h}^{+}) & = & \hat{\Delta}_{u}(t_{h})-\alpha e_{u}(t_{h}).\label{eq:delta-pi}
\end{eqnarray}
Therefore, the update equations of PISync algorithm can be described
with the following matrix equation:

\begin{eqnarray}
\underset{X(h+1)}{\underbrace{\left[\!\!\!\begin{array}{c}
e(h+1)\\
\hat{\Delta}_{u}(h+1)
\end{array}\!\!\!\right]}} & \!\!\!\!\!\!\!\!=\!\!\!\!\!\!\!\! & \left[\!\!\!\begin{array}{cc}
0 & \int_{t_{h}}^{t_{h+1}}f_{u}(t)dt\\
0 & 1-\alpha\int_{t_{h}}^{t_{h+1}}f_{u}(t)dt
\end{array}\!\!\!\right]\underset{X(h)}{\underbrace{\left[\begin{array}{c}
e(h)\\
\hat{\Delta}_{u}(h)
\end{array}\right]}}\nonumber \\
 &  & +(B+\mathcal{T}_{h+1}-\mathcal{T}_{h})\!\!\left[\!\!\!\!\begin{array}{c}
-1\\
\alpha
\end{array}\!\!\!\!\right].\label{eq:state_space-pi}
\end{eqnarray}
Taking the expectation of both sides yields 

\begin{eqnarray}
E\!\left\llbracket X(h+1)\right\rrbracket \!\!\!\!\!\! & = & \!\!\!\!\!\!\!\underset{A}{\underbrace{\left[\!\!\!\!\begin{array}{cc}
0 & Bf_{0}\\
0 & \!\!\!\!\!1-\alpha Bf_{0}
\end{array}\!\!\!\!\right]}\!}E\!\left\llbracket X(h)\right\rrbracket \!+\!\!\left[\!\!\!\begin{array}{c}
-B\\
\alpha B
\end{array}\!\!\!\right].
\end{eqnarray}
and the eigenvalues of the matrix $A$ can be obtained as:

\begin{align}
\lambda_{1}=0, & \lambda_{2}=1-\alpha Bf_{0}.\label{eq:roots-pi}
\end{align}
Consequently, asymptotic convergence is established, if and only if

\begin{align}
0 & <\alpha<\frac{2}{Bf_{0}}.\label{eq:step_size_bounds-pi}
\end{align}
Finally, we get for $\hat{\Delta}_{u}(\infty)$ and $e_{u}(\infty)$
that 

\begin{eqnarray}
\hat{\Delta}_{u}(\infty) & = & \left(1-\alpha Bf_{0}\right)\hat{\Delta}_{u}(\infty)+\alpha B=\frac{1}{f_{0}},\\
e_{u}(\infty) & = & B(\hat{\Delta}_{u}(\infty)f_{0}-1)=0.
\end{eqnarray}

\subsection{Asymptotic Variance}

Following the steps in Appendix-\eqref{appendix:Asymptotic-Variance},
$\hat{\Delta}(h+1)$ can be written as:

\begin{eqnarray}
\hat{\Delta}(h+1) & = & \hat{\Delta}(h)-\alpha e(h+1)\nonumber \\
 & = & \frac{z_{h}+1}{f_{0}}\!-\alpha\!\left(\!\!z_{h}\!\!\left(\!\!B\!\!+\!\frac{w_{h+1}}{f_{0}}\!\!\right)\!\!+\!\frac{w_{h+1}}{f_{0}}\!\!-\!d_{h+1}\!\!\right)\nonumber \\
 & = & z_{h}\left(\frac{1}{f_{0}}-\alpha\left(B+\frac{w_{h+1}}{f_{0}}\right)\right)+\frac{1}{f_{0}}\nonumber \\
 &  & -\alpha\left(\!\!\frac{w_{h+1}}{f_{0}}\!\!-\!d_{h+1}\!\!\right).
\end{eqnarray}
Then, $z_{h+1}=\hat{\Delta}(h+1)f_{0}-1$ becomes:

\begin{eqnarray}
z_{h+1} & = & z_{h}\left(\!1-\alpha g_{h+1}\!\right)\!-\!\alpha g_{h+1}\!+\!\alpha f_{0}(B+d_{h+1}).
\end{eqnarray}
After some straightforward steps, we obtain the asymptotic value of
$E\left\llbracket z_{h}\right\rrbracket $ as

\begin{eqnarray}
\underset{h\rightarrow\infty}{lim}E\left\llbracket z_{h}\right\rrbracket  & = & 0
\end{eqnarray}
and the asymptotic value of $E\left\llbracket z_{h}^{2}\right\rrbracket $
as

\begin{eqnarray}
\underset{h\rightarrow\infty}{lim}E\left\llbracket z_{h}^{2}\right\rrbracket  & = & \frac{\alpha\left(\frac{Bf_{max}^{2}}{3}+f_{0}^{2}\sigma_{d}^{2}\right)}{2Bf_{0}-\alpha\left(B^{2}f_{0}^{2}+\frac{Bf_{max}^{2}}{3}\right)}.
\end{eqnarray}
Finally, we calculate mean squared-error, which is also the asymptotic
variance as:

\begin{eqnarray}
E\left\llbracket \left(e(h+1)\right)^{2}\right\rrbracket  & = & E\left\llbracket z_{h}^{2}\right\rrbracket \left(B^{2}+\frac{E\left\llbracket w_{h+1}^{2}\right\rrbracket }{f_{0}^{2}}\right)\nonumber \\
 & = & +\frac{E\left\llbracket w_{h+1}^{2}\right\rrbracket }{f_{0}^{2}}+E\left\llbracket d_{h+1}^{2}\right\rrbracket \nonumber \\
 & =  & \frac{\alpha\left(\!\frac{Bf_{max}^{2}}{3}\!+\!f_{0}^{2}\sigma_{d}^{2}\!\right)}{2Bf_{0}\!-\!\alpha\!\left(\!\!B^{2}f_{0}^{2}\!+\!\frac{Bf_{max}^{2}}{3}\!\!\right)}\!\!\left(\!\!B^{2}\!+\!\frac{Bf_{max}^{2}}{3f_{0}^{2}}\!\!\right)\nonumber \\
 &  & +\frac{Bf_{max}^{2}}{3f_{0}^{2}}+\sigma_{d}^{2}.\label{eq:pisync_variance}
\end{eqnarray}

\end{document}